\documentclass[english,a4paper,twocolumn,longbibliography]{revtex4-1}
\usepackage[T1]{fontenc}
\usepackage[latin9]{inputenc}
\usepackage{url}
\usepackage[unicode]{hyperref}
\hypersetup{
   unicode=true,          
   plainpages=false,
   colorlinks=true,       
   citecolor=blue,        
}
\makeatletter
\newcommand{\lyxaddress}[1]{
\par {\raggedright #1
\vspace{1.4em}
\noindent\par}
}

\makeatother

\usepackage{babel}
\begin{document}

\title{Comment on: `Single-shot simulations of dynamic quantum many-body
systems'}

\author{P. D. Drummond$^{1}$ and J. Brand$^{2}$}
\maketitle

\lyxaddress{$^{1}$Centre for Quantum and Optical Science, Swinburne University
of Technology, Melbourne, Victoria 3122, Australia\\
$^{2}$Dodd-Walls Centre for Photonics and Quantum Technology, Centre
for Theoretical Chemistry and Physics, New Zealand Institute for Advanced
Study, Massey University, Private Bag 102904 North Shore, Auckland
0745, New Zealand}

In their recent paper\cite{sakmann2016single}, Sakmann and Kasevich
study the formation of fringe patterns in ultra-cold Bose gases and
claim: `Here, we show how single shots can be simulated from numerical
solutions of the time-dependent many-body Schrödinger equation.' It
would be remarkable if they had solved this exponentially complex
equation. Instead they solve nonlinear equations with the aim to approximate
the solution of the Schrödinger equation.  The authors proceed to
criticize phase-space approaches to simulating quantum dynamics and
claim the impossibility of interpreting single trajectories of the
truncated Wigner (tW) method as single-shot experimental outcomes.
Here we aim to provide relevant context and elaborate why we disagree
with the authors' claims.

The multi-configurational time-dependent Hartree or MCTDHB equations
used by the authors are based on a multi-mode expansion\cite{Alon:2008}
that is conjectured to converge to the solutions of the Schrödinger
equation when many terms are taken into account. Truncating the expansion
with a small number $M$ of modes limits the exponential complexity
and makes the problem solvable on a computer. Figure 3  demonstrates
the approximate nature of this procedure with $N=10$ particles but
does not vindicate the truncation applied for the other two examples,
where different observables are appoximated with fewer modes and larger
particle numbers. For the collision of attractive condensates with
$N=100$ particles the expansion is truncated after $M=4$ modes (Fig.~1)
and for the fluctuating vortices after $M=2$ (Fig.~2, $N=10000$).
We do not and cannot know how well the MCTDHB expansion approximates
the true solution of the Schrödinger equation in the many-particle
regime because reliable convergence criteria are only applicable in
very specific cases, convergence control is more difficult for larger
particle numbers\cite{Cosme2016}, and proposed error estimation procedures
have not been implemented yet\cite{Lee2014}. Specifically, the free-space
dynamics of attractive condensates was recently demonstrated to have
a slowly-convergent MCTDHB expansion\cite{Cosme2016}. It is clear that
the MCTDHB wave function is, at best, approximate. 

The authors criticize the tW approximation for a subtle reason: For
pure states, only Gaussians give exact, positive Wigner distributions.
Since the tW distribution is not Gaussian and is positive, so the
authors argue, it cannot represent a pure quantum state and hence
the single-shot interpretation of Wigner phase space trajectories
has to be dismissed.

The flaw in this argument is that, like the MCTDHB approach, the tW
approximation is not exact but an approximation. Approximate solutions
cannot be expected to obey every theorem holding for the exact result
and solutions of the truncated MCTDB equations do not do this either,
e.g.~with respect to the separation of relative and centre-of-mass
motion\cite{Cosme2016}. The exact Wigner function contains the same
information as the wave function. The tW approximation includes the
first quantum corrections of a ($1/N$) expansion, and neglects terms
of $\mathcal{O}(1/N^{2})$, which can have negative values. Despite
this, the tW method\cite{Steel1998} is well tested against exact
methods and experiment\cite{Corney_2008}. It gives small errors for
quantum correlations when $N$ (the number of particle per mode) is
large, provided time-scales are not too long\cite{Sinatra_2002_truncatedWinger}.

The Wigner method is known to have marginal distributions that are
the probability of quadrature measurements, and this property is independent
of the positivity of the total distribution\cite{Hillery_Review_1984_DistributionFunctions}.
Because of this, it is ideally suited to single-shot interpretations,
provided the corresponding measurements are used. Particle instead
of quadrature measurements can be interpreted as approximating single
shots in volumes with large mode occupation\cite{lewis2016approximate},
or discrete types of number-phase Wigner distributions can be constructed\cite{Hush:2010}.

In summary, the authors  have only solved the Schrödinger equation
with an approximation whose accuracy is unclear. Criticising other
approximate approaches as not being exact is hypocritical. A separate
question concerns the scientific value of simulating experimental
single shots using pure quantum states. By definition they are not
reproducible. Quantitative tests of theories can only be performed
on reproducible results, e.g.~on correlation functions that are obtained
by averaging over many observations. In current experiments, these
usually involve impure density matrices with finite temperature and
particle number variation.

%

\end{document}